\documentclass{article}

    \usepackage[nonatbib, preprint]{neurips_2021}

\usepackage[utf8]{inputenc} 
\usepackage[T1]{fontenc}    
\usepackage{hyperref}       
\usepackage{url}            
\usepackage{booktabs}       
\usepackage{amsfonts}       
\usepackage{nicefrac}       
\usepackage{microtype}      
\usepackage{xcolor}         

\usepackage{amsmath,amssymb, bbm}
\usepackage{algorithm}
\usepackage[english]{babel}
\usepackage{cleveref}
\usepackage{graphicx}
\usepackage{caption}
\usepackage{subcaption}
\usepackage{multirow}

\usepackage[backend=bibtex,style=ieee]{biblatex}
\title{Reinforcement Learning for Industrial Control Network Cyber Security Orchestration}

\author{%
  John Mern\thanks{PhD Candidate, Stanford Intelligent Systems Laboratory} \\
  Department of Aeronautics and Astronautics\\
  Stanford University\\
  Stanford, CA 94305 \\
  \texttt{jmern91@stanford.edu} \\
  \And
  Kyle Hatch\thanks{Undergraduate Student, Stanford University} \\
  Department of Computer Science\\
  Stanford University\\
  Stanford, CA 94305 \\
  \texttt{khatch@stanford.edu} \\
  \And
  Ryan Silva  \\ 
  Applied Physics Laboratory\\
  Johns Hopkins University\\
  Laurel, MD 20723 \\
  \And
  Jeff Brush  \\ 
  Applied Physics Laboratory\\
  Johns Hopkins University\\
  Laurel, MD 20723 \\
  \And
  Mykel J. Kochenderfer\thanks{Associate Professor, Stanford Intelligent Systems Laboratory} \\
  Department of Aeronautics and Astronautics\\
  Stanford University\\
  Stanford, CA 94305 \\
  \texttt{mykel@stanford.edu} \\
}

\begin{document}

\maketitle

\begin{abstract}
    Defending computer networks from cyber attack requires coordinating actions across multiple nodes based on imperfect indicators of compromise while minimizing disruptions to network operations.
    Advanced attacks can progress with few observable signals over several months before execution. 
    The resulting sequential decision problem has large observation and action spaces and a long time-horizon, making it difficult to solve with existing methods. 
    In this work, we present techniques to scale deep reinforcement learning to solve the cyber security orchestration problem for large industrial control networks. 
    We propose a novel attention-based neural architecture with size complexity that is invariant to the size of the network under protection.
    A pre-training curriculum is presented to overcome early exploration difficulty. 
    Experiments show in that the proposed approaches greatly improve both the learning sample complexity and converged policy performance over baseline methods in simulation.
\end{abstract}

\section{Introduction}
Cyber attacks have been increasingly targeting computer networks that are integrated with industrial control systems (ICS)~\cite{das2020}.
Attack techniques focusing on stealth and redundant access make them difficult to detect and stop~\cite{alladi2020}. 
Intrusion detection systems monitor networks and alert security orchestrators to potential intrusions, though  high false-alarm rates cause many alerts to receive no security response. 
Advanced persistent threat (APT) attackers take advantage of this by spreading across target networks undetected over long periods to prepare an eventual attack~\cite{li2016}.
In addition to the theft of sensitive data common in APT attacks on standard networks, attacks on ICS can additionally result in disruption of the controlled system, physical destruction of equipment, and even loss of life~\cite{langner2011, liang2016, di2018}.

It is not feasible for human security teams to respond to every potential intrusion alert and to reliably mitigate all threats.
This work seeks to demonstrate the feasibility of developing an automated cyber security orchestrator (ACSO) to assist human analysts by automatically investigating and mitigating potential attacks on ICS networks. 
We first present a model of this task as a discrete-time sequential decision making problem. 
The proposed model captures many of the challenges of the underlying decision problem while remaining agnostic to specific network configuration and communication dynamics.
To support this, we implemented a cyber attack simulator that can efficiently generate large quantities of trial data.

Many sequential decision methods cannot solve this problem because explicit models of APT attack dynamics and alert behaviors are not generally known.
Reinforcement learning (RL) may be used to train policies for complex tasks without explicit models~\cite{vinyals2019, hoel2019}, however the ACSO problem poses several challenges to existing reinforcement learning (RL) solvers.
The observation and action spaces grow with the number of nodes on the protected network, leading to very high-dimensional, discrete observation and action spaces.
Deep reinforcement learning is known to struggle to learn over large, discrete action spaces due to the difficulty of generalizing over actions~\cite{dulac2015}.
Similarly, large input spaces have been found to decrease the sample efficiency of stochastic learning~\cite{nguyen2020}.

APT attacks are designed to be difficult to detect, causing limited observability of the true compromise state of nodes on the network. 
Partially observable problems are known to be significantly more difficult to solve than their fully observable counterparts, requiring learning over sequences of observations to infer unseen states~\cite{hausknecht2015}.
Effective defense against APTs requires timely response to compromises, though intrusion campaigns can last several months.
Modeling the decision process requires fine resolution time-steps, leading to very long time horizons.
Long time-horizons and sparse rewards can drastically increase the sample complexity of learning, both through exploration difficulty and temporally delayed credit assignment~\cite{andrychowicz2017, arjona-medina2019}.

This paper proposes a neural network architecture and training curriculum that scales to the large problem space.
We present an attention-based architecture that effectively learns over many cyber network nodes without growth in the number of required parameters. 
The lower parameter complexity network is shown to significantly accelerate training while converging to a higher performing solution than a larger baseline architecture. 
To overcome exploration difficulty, we introduce a supervised pretraining step that minimizes a composite large-margin classification and temporal-difference loss over a set of expert trajectories. 

We tested the proposed solution methods against baseline neural network architectures and training approaches, as well as against expert-designed policies. 
These experiments demonstrate how each component contributes to the proposed method's scalability and suggest that deep RL is a viable method to implement an autonomous network security agent. 
The methods presented may be generalized to scale RL to other problems with large input spaces, output spaces, or long time horizons.

\section{Background}
\subsection{Reinforcement Learning}
In sequential decision problems, the environment is modeled by states $s$ that evolve according to potentially stochastic dynamics. 
An agent takes actions $a$ that condition state transition distributions $T(s'\mid s, a)$ and generate rewards $r(s,a,s')$. 
In many problems, the state of the world is not known. 
In these partially observable domains, agents receive noisy observations according to $o \sim Z(o \mid s, a)$.


A sequential decision problem is solved by an action sequence that maximizes the expected value $ V(s) = \mathbb{E}\big[ \sum_t \gamma^t r(s_t, a_t)\big]$ for all states in the trajectory.  
Reinforcement learning methods learn a policy $\pi : o_{t-\tau:t} \mapsto a_t$ that maps a history of observations to actions through repeated trial-and-error with the environment. 
In each trial, actions are taken according to the current policy until a terminal condition is reached. 
At the end of the trial, the policy is updated to improve the expected performance.

Reinforcement learning methods that use neural networks to represent the learned policy are known as \emph{deep} reinforcement learning.
Deep Q-Network (DQN) learning is a popular method in which the learned neural network predicts the expected value $Q(o_{t-\tau:t}, a) = \mathbb{E}\big[r(s_t, a) + \gamma V(s_{t+1})\big]$ of taking each action in the action space for a given input history~\cite{mnih2013}. 
The Q-network is used as a policy by taking the action with the highest predicted value $a^* = \mathrm{argmax}_a Q(o_{t-\tau:t}, a)$.

Deep RL training typically requires large amounts of data~\cite{wei2019}. 
The amount of trials required to solve a task tends to grow with the task complexity and neural network size, with more complex tasks often requiring larger neural networks. 
Reinforcement learning agents improve by stochastically exploring new trajectories with each trial.
Tasks with very large input spaces, output spaces, or very long time horizons tend to require much more exploration because the odds of finding a successful trajectory through random exploration are low~\cite{ecoffet2021}. 
\subsection{Networked Industrial Control Systems}
Industrial control systems are networks of devices used to monitor and control a coordinated physical process, such as an assembly line or power plant.
Simple operational technology computing devices such as programmable logic controllers (PLCs) are installed on the industrial equipment to receive sensor readings or to issue control commands.
These PLCs are networked with information technology networks to enable remote access and control~\cite{stouffer2011}.

Integration of ICS networks with internet-connected networks has made them vulnerable to cyber attack~\cite{glenn2016}.
Despite this increased risk, networking is often an operational requirement for processes where system components are difficult to access or geographically distant.
For example, electrical power grid management requires coordination of substations that may be distributed over very large areas.
To attempt to mitigate this vulnerability, many networks are organized into firewall-separated levels, with low-privilege nodes on less restricted and internet-accessible levels, and process-critical nodes on more isolated subnets~\cite{stouffer2011}.

\begin{figure}
    \centering
    \includegraphics[width=1.0\columnwidth]{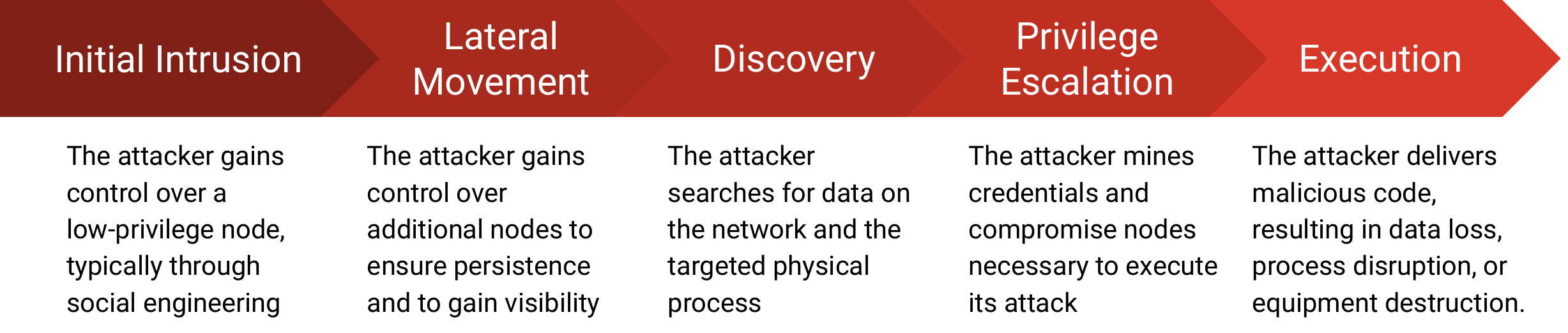}
    \caption{APT Attack Progression: This shows the typical progression of an APT attack at the level of tactics in the MITRE ATT\&CK framework. The process starts on the left with initial intrusion, proceeding over several months to eventual execution.}
    \label{fig:kill_chain}
\end{figure}

Many recent, high-profile infrastructure attacks were executed by advanced persistent threats (APTs).
Advanced persistent threats comprise the top two tiers of the Defense Science Board threat taxonomy and are typically well-funded and willing to spend significant time to achieve their goal~\cite{gosler2013}. 
APTs overcome the layered ICS network structure by first compromising a low-privilege node in the target network~\cite{meckl2017}. 
Using this node, the APT will then conduct internal network reconnaissance and take control over more privileged nodes until it gains the authority needed to achieve its primary objective, for example to destroy equipment. 
A typical APT attack process, based on the tactics of the MITRE ATT\&CK framework~\cite{strom2016}, is shown in~\cref{fig:kill_chain}.

During reconnaissance and lateral movement, APT activity tends to be very difficult to detect with automated intrusion detection systems~\cite{li2016}. 
The lateral movement phase also comprises the majority of the time of the attack campaign, often taking several months to complete. 
Once an attack is staged, however, it can usually be executed in a matter of hours. 
Because of this, successfully defending against an APT attack requires detecting and securing the compromised nodes in the staging phases. 



Machine learning methods have been applied to many aspects of cyber security, most extensively to perception tasks such as malware recognition and anomaly detection~\cite{apruzzese2018, meckl2017, milajerdi2019, naseer2018}.
Reinforcement learning has been applied to cyber security in various capacities. 
\textcite{nguyen2019} provide a survey of recent work in this area. 
The scope of each work referenced in this survey is limited in the type of network under protection, the actions available to the defender agent, and/or to the type of attack against which it is defending. 
For example, \textcite{gupta2018} propose a system restricted to filtering signals spoofing networked sensor measurements.

\section{Problem Formulation}
\begin{figure}[tb]
    \centering
    \includegraphics[width=1.0\columnwidth]{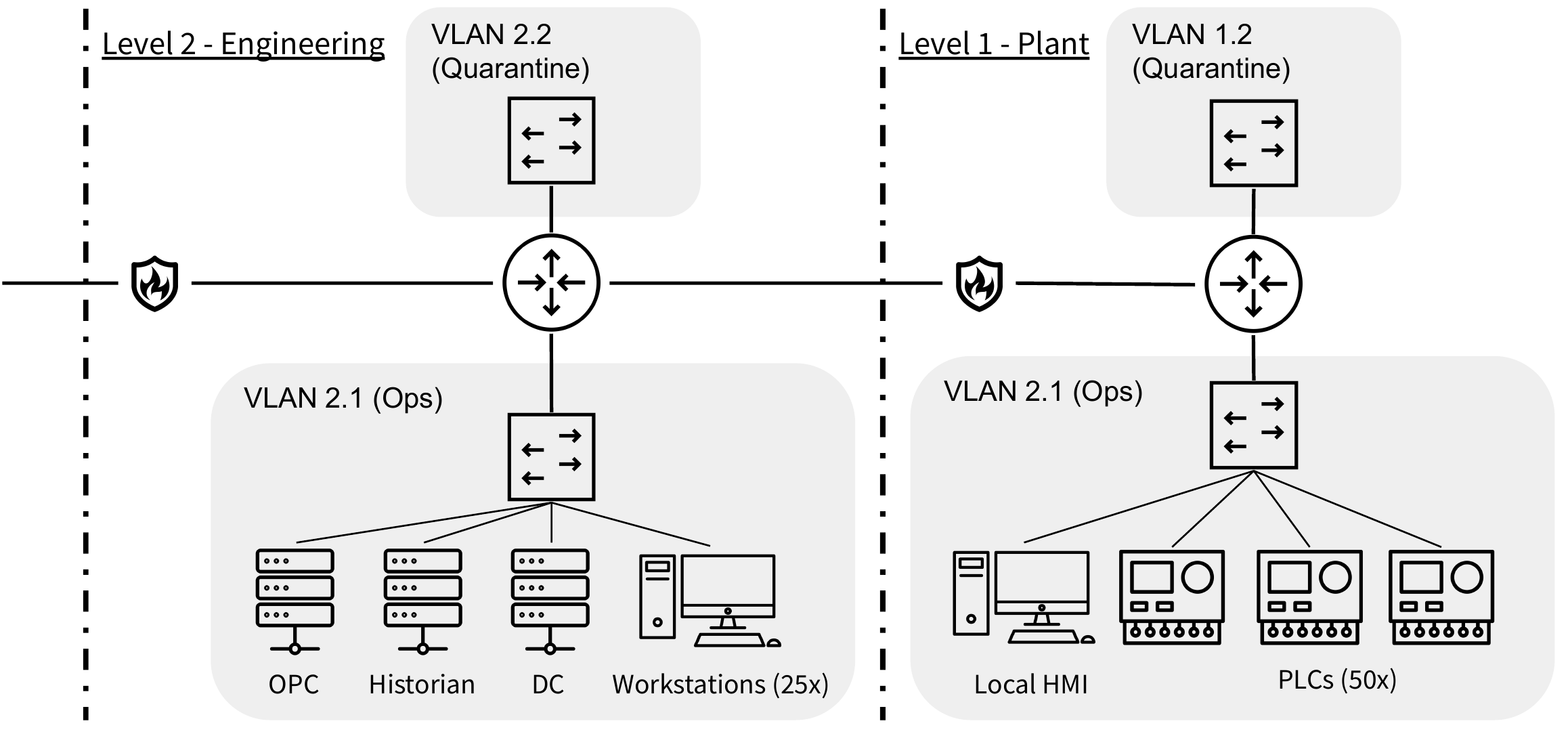}
    \caption{Simulated Network Architecture: The problem simulates level 2 and level 1 of the PERA model. Each level contains an operations VLAN and a nominally empty quarantine VLAN. Level 2 contains twenty-five workstation nodes and three servers. Level 1 has five local human-machine interface nodes and fifty networked PLCs. All network message traffic is simulated through virtual switches, routers, and firewalls. The computing node for the ACSO is not explicitly modeled.}
    \label{fig:ics}
\end{figure}

The objective is to prevent an APT from disrupting a set of PLCs while minimizing how much the defensive actions interfere with network operations.
This work is restricted to defending the engineering level (2) and plant level (1) of the Purdue enterprise reference architecture (PERA)~\cite{williams1993} as shown in~\cref{fig:ics}.
Level 2 contains exclusively workstations and servers, and level 1 contains workstations and all PLCs.
The network studied  in this work has twenty-five workstations in level 2, five workstations in level 1, and fifty PLCs for a total of eighty-three nodes. 

The state space for the problem is the joint space of the ICS network states and the APT attacker states. 
The ICS network state is the joint state of all nodes on the network. 
Server and workstation node states define how much access and control the APT has and where on the network the node is located. 
A PLC state defines whether or not it has been disrupted by an attack. 
The APT state defines the current APT sub-goal, knowledge, and available resources. 

The ACSO agent has full knowledge of the network configuration and PLC states, but observes node compromise state through alerts. 
Alerts are actively generated by investigative actions taken by the ACSO, which may generate false negatives, and passively by an intrusion detection system triggered by APT actions on compromised nodes with many false alarms.
Each episode begins with a single level 2 workstation compromised by an APT.
The episode terminates when 25 PLCs have been disrupted by APT attacks or after 5,000 hours have been simulated. 
Each decision step is one hour of simulation world time and the discount factor is 0.999.

Each step, the ACSO can choose to take an action on a single node or do nothing.
There are seven valid action types for workstation nodes, six action types for server nodes, and two action types for PLCs. 
A total of 329 unique actions may be taken across the nodes.
With every step, the APT also takes actions based on its state and a stochastic attacker model. 
Actions are not necessarily instantaneous and some actions will complete over several time-steps. 
During this time, agents can initiate additional actions. 

Reward each step is based on the fraction of PLCs that are operating nominally and the inconvenience caused by the ACSO actions taken. 
The reward function is defined as
\begin{align}~\label{eq: reward}
    r(s,a) &= r_\text{PLC}(s, a) + r_\text{IT}(s, a) + r_\text{Term}(s,a) \\
    &= \Big(1 - 0.04\sum_{p \in \text{PLCs}} 1\{p \mathrm{\ disrupted}\}\Big) \Big(1.0 - \sum_{a \in A_t}\mathrm{cost}(a)\Big) + \frac{1}{\gamma}1\{s_\text{time} \geq t_\text{max}\}
\end{align}
where $A_t$ is the set of all actions completing at time step $t$.

The first term scales reward by the number disrupted PLCs, where disruption of twenty five PLCs results in process shutdown.
The second term imposes a penalty for ACSO actions, where each action is assigned a cost based on its perceived burden to network operations.
Action costs were elicited from network and security experts.
For example the low-disruption action of rebooting a workstation has a cost of 0.01, while the more disruptive action of re-imaging a server has a cost of 0.05.
The final term rewards the agent for reaching the episode time limit $t_{\max}$.
The $1/\gamma$ magnitude of the terminal reward ensures the optimal state value does not drift with episode time. 

To facilitate this and future research, we developed an ICS network attack simulation (INASIM) in the Julia language.
To the authors knowledge, this is the first open-source APT attack simulation capable of generating data in quantities required for deep reinforcement learning on a single workstation. 
INASIM implements a configurable network simulation and a baseline APT attacker model, as well as an API for custom attacker agents.
The simulation is compatible with the POMDPs.jl~\cite{Egorov2017} framework as well as with the OpenAI Gym~\cite{brockman2016} framework through supplied Python bindings.
The simulation and baseline attacker model are described briefly here. 
A description of states, actions, and environment dynamics is provided in the appendix along with numerical settings used for the studies in this work. 

The network simulation defines the structure of the network, the state of the nodes, and the actions available to the APT and ACSO agents.
The ACSO can take actions to investigate node states or to secure compromised nodes.
The APT can take actions to gain further control over a node it has already compromised, or to use a compromised node to find and infect additional nodes.
The APT can only act from nodes it has already compromised, though the ACSO can act on any node in the network. 
Each node's state defines how much it has been compromised by the APT. 
Different types of compromise enable the APT to take different actions from or on that node. 
Compromise levels also change the way the defensive ACSO actions affect the node.
For example, an APT can take actions on a node in the ``escalated privileges'' state to resist ACSO actions to change the password. 


The baseline APT agent is modeled as a stochastic finite state machine. 
The machine states correspond to the attack-phase tactics shown in~\cref{fig:kill_chain}.
Every time-step, the APT first updates its machine state based on its belief over the network. 
Each machine-state defines a set of exit-criteria and a stochastic rules-based sub-policy.
The APT may have one of two goals and may use one of two attack vectors for each episode. 
The APT goal may be to disrupt the PLC process under control or to destroy the PLC-controlled equipment. 


\section{Solution Methods}
The proposed solution uses an augmented DQN algorithm to learn a policy defined by our attention-based neural network.
We implemented several extensions to the baseline algorithm, based on studies of Rainbow DQN~\cite{hessel2018}.
The included extensions are double-DQN~\cite{hasselt2010}, prioritized experience replay~\cite{schaul2016}, and $n$-step TD loss~\cite{sutton1998}.
The training loss for a given step is 
\begin{equation}~\label{eq: td_loss}
    \mathcal{L}_{TD}\big(Q_\pi(h_t, a)\big) = \Big\| \Big(\sum_{\tau=t+1}^{t+n} \gamma^{\tau - t}r_\tau + \gamma^nQ_\phi\big(h_{t+n}, \mathrm{argmax}_{a'} Q_\pi(h_{t+n}, a')\big)\Big) - Q_\pi(h_t, a) \Big\|
\end{equation}
where $h_t$ is the history of observations at time $t$, $Q_\pi$ is the action value estimate of the policy network and $Q_\phi$ is the action value estimate of the target network. 
The $\| \cdot \|$ represents the Huber-loss norm. 
This loss is calculated over batches of importance-weighted samples from an experience replay buffer and used to estimate the gradient for each network update. 

In addition to the task reward presented in~\cref{eq: reward}, a shaping reward was defined based on the formulation of~\textcite{ng1999}.
This reward was designed to incentivize the agent to secure compromised nodes. 
The shaping function was defined as  
\begin{equation}~\label{eq: train_reward}
    r_{shape}(s,a,s') = \gamma(A\delta_{W} + B\delta_{S})
\end{equation}
where $\delta_{W}$ and $\delta_{S}$ are changes in the number of workstations and servers compromised by the APT from state $s$ to $s'$, respectively, and $A$ and $B$ are weight factors.
The weighted sum of~\cref{eq: reward} and~\cref{eq: train_reward} were used for training.
Only~\cref{eq: reward} was used for evaluation.

Training hyper-parameters were tuned by a grid search training on a smaller ICS network with ten level 2 workstation nodes, three level 1 workstations nodes, and thirty PLCs. 
We searched over the shaping reward weight, the observation-history interval, the target network update frequency, and the $\epsilon$-greedy exploration decay schedule. 
The parameter set leading to the highest average performing policy over several seeds after 500 episodes was selected. 
We used the PyTorch framework for neural network implementation and training~\cite{paszke2015}.
Numerical values for the training parameters as well as additional practical details on the training process can be found in the appendix.

\subsection{Neural Network}~\label{sec: neural_net}
\begin{figure}[tb]
    \centering
    \includegraphics[width=0.95\columnwidth]{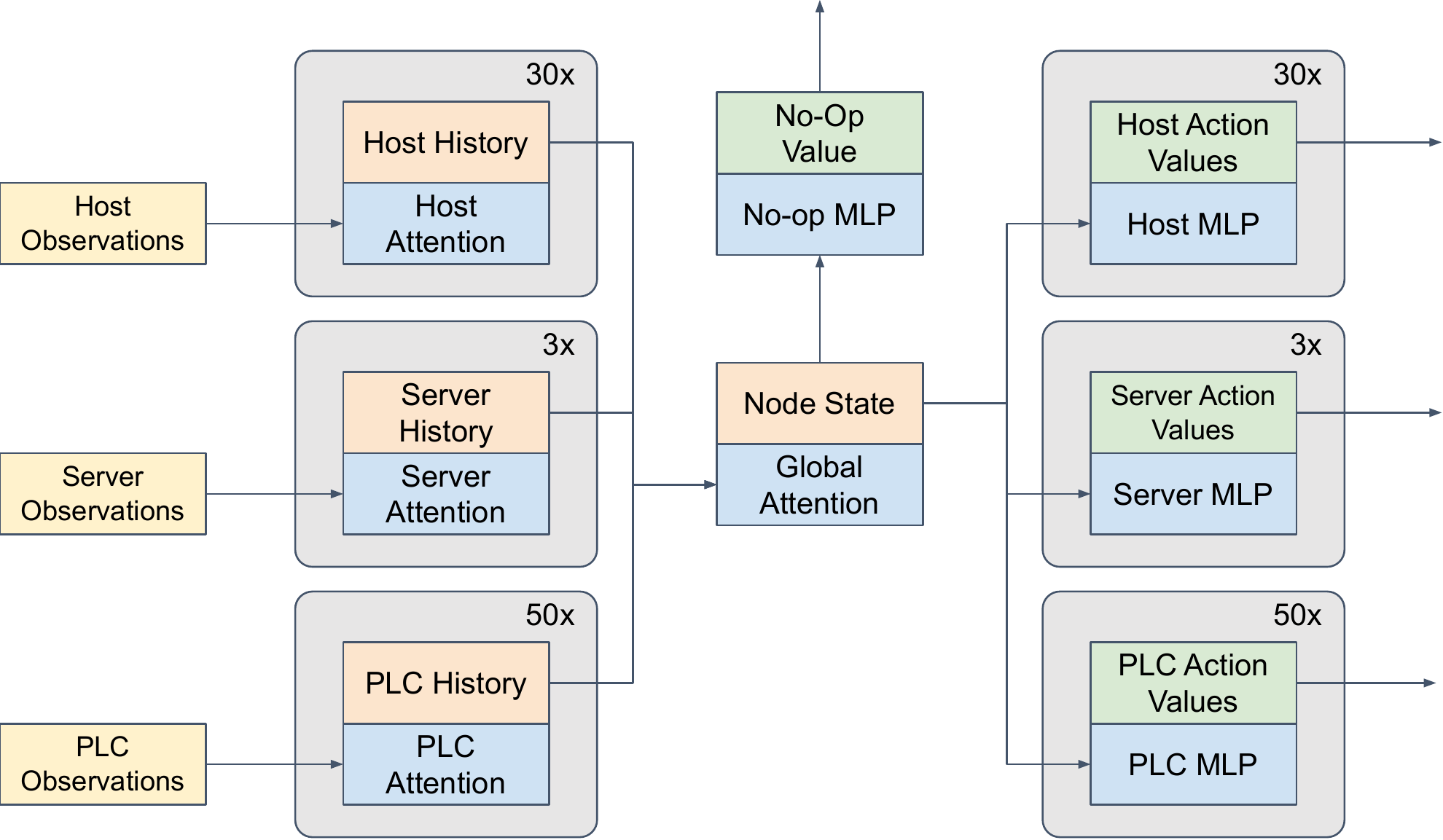}
    \caption{Attention Network: Inputs for each network node are passed into attention sub-graphs based on node type. These sub-graphs embed each node's history to a latent vector representation. These vectors are stacked and passed through a self attention sub-graph, to provide global context to each latent vector. These vectors are then passed through fully connected sub-graphs to output action value estimates for each node.}
    \label{fig:neural_network}
\end{figure}
The ACSO problem has input and output spaces with dimensions that scale with the number of nodes in the ICS network.
Node observations are represented as vectors $o^{(i)}_t$, where each element of the vector indicates whether the corresponding alert or action was observed on that node $i$ during that step $t$.
Workstation node observations have 16 elements, server node observations have 14 elements, and PLC observations have 7 elements. 
The complete observation for a given time-step contains the observation vector for each node in the ICS network, with 792 total elements. 

Because the problem is partially observable, a history of prior observations was used as the network input each step. 
At time $t$, a history of the previous $\tau$ steps was input to the network, such that the complete input was $H_t = O_{t-\tau:t}$.
In this work, 256 prior steps were used, for a total input size of 202,752.
The output space consists of all valid actions at each time-step, for a total of 329 elements. 

Specialized neural architectures can take advantage of structure inherent in input data to reduce complexity.
For example, convolutional nets leverage spatial invariance in images and recurrent networks encode correlation in sequences~\cite{krizhevsky2012, cho2014}.
We designed the neural network shown in~\cref{fig:neural_network} using attention mechanisms~\cite{vaswani2017} to accommodate the large input space without significant degradation of explanatory capacity or intractable growth in size. 
Attention mechanisms have been shown to improve learning efficiency on tasks with exchangeable inputs~\cite{mern2019, mern2020}.

Each node history is input to its own sub-graph which embeds its history array to a single latent state vector.
All sub-graphs of a given node type share the same parameter set, so a growth in the number of nodes does not cause a growth in the total number of network parameters. 
The latent states of each individual node are then stacked and input to a global attention sub-graph. 
This sub-graph allows the network to learn which features of neighboring nodes are relevant to the value function of actions on a given node. 
These contextualized node vectors are then passed to feed-forward output sub-graphs. 
Like the input sub-graphs, network parameters are shared between nodes of the same type. 

\begin{figure}[tb]
    \centering
    \begin{subfigure}[b]{0.35\columnwidth}
        \centering
        \includegraphics[width=\columnwidth]{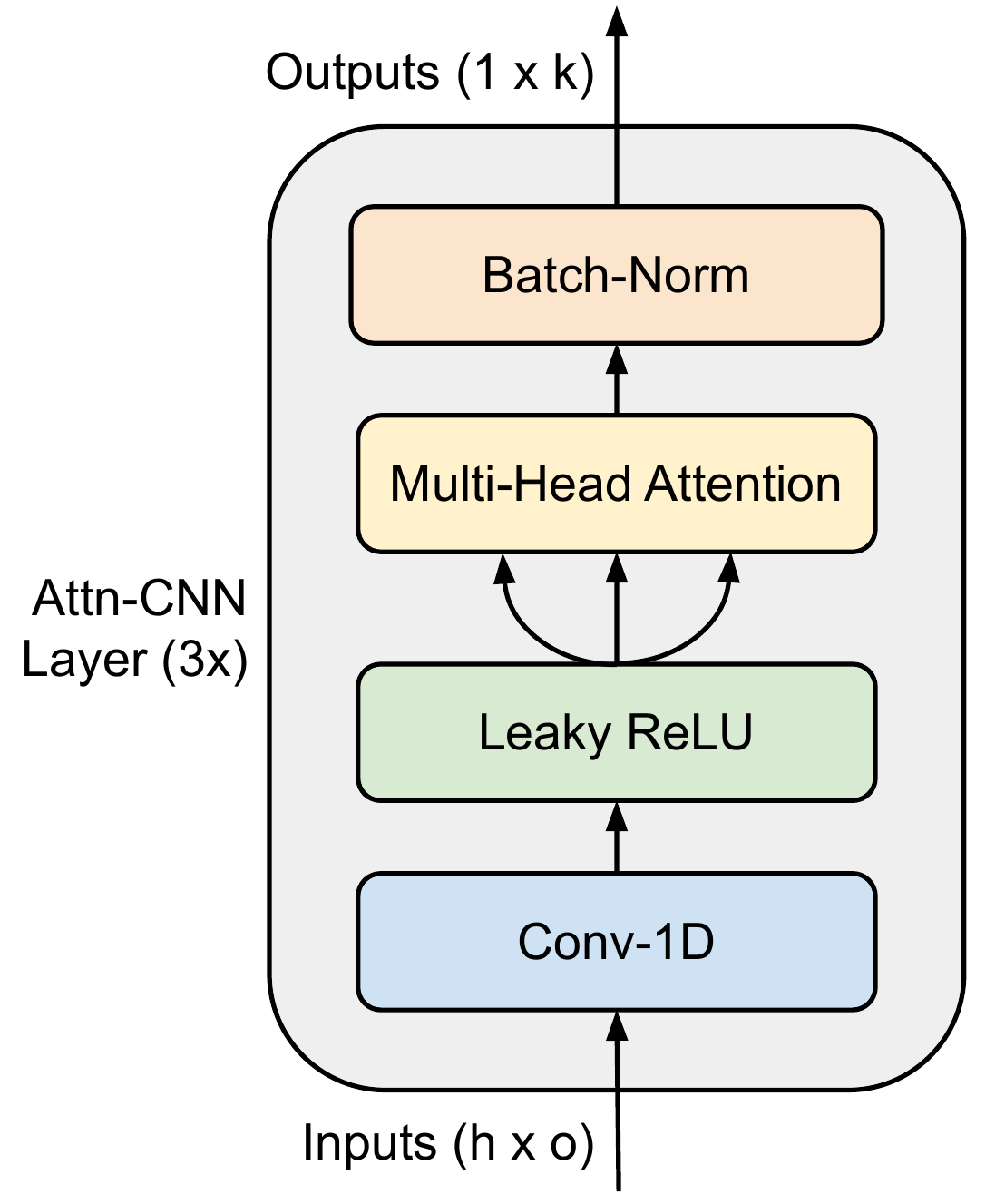}
        \caption{Temporal Attention}
    \end{subfigure}
    \hfill
    \begin{subfigure}[b]{0.55\columnwidth}
        \centering
        \includegraphics[width=\columnwidth]{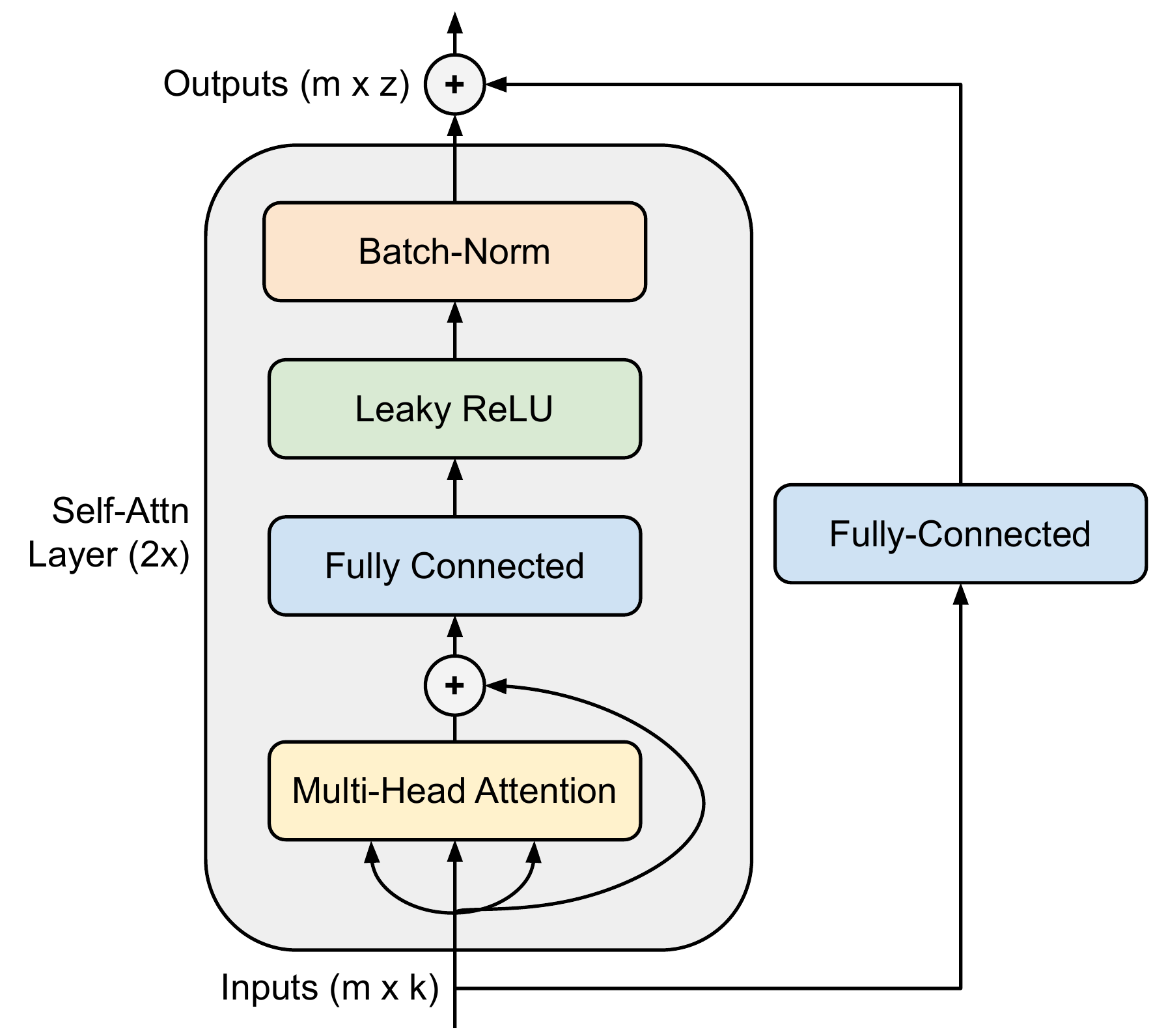}
        \caption{Global Attention}
    \end{subfigure}
    \caption{Attention Sub-Graphs: Figure (a) shows the temporal attention graph, which alternates layers of multi-headed attention with 1D temporal convolution. Figure (b) shows the global attention sub-graph, which alternates fully connected layers with multi-headed attention. A skip connection is included to facilitate gradient back-propagation.}
    \label{fig:Attn mechs}
\end{figure}

The temporal attention and global attention sub-graphs are shown in~\cref{fig:Attn mechs}.
The temporal attention sub-graphs alternate 1D convolution in the time dimension with multi-headed, dot-product attention layers so that the network may learn to attend to elements of the input sequence that are important to solving the problem. 
The global self-attention sub-graph alternates multi-headed dot-product attention mechanisms with affine layers. 
This structure allows the output representation corresponding to a node $i$ to include relevant information from any other node in the network. 
Including this learned global context in the latent representation of each node allows the output space to be factored to individual sub-graphs for each node. 

The proposed neural network architecture had a total of 683,108 parameters. 
This size does not necessarily vary with the number of ICS nodes considered. 
A convolutional neural network was implemented as a baseline for comparison. 
The baseline network had four 1D temporal convolutional layers, followed by a fully connected layer. 
The number of parameters of the convolutional network grows with the with the number of network nodes nodes, with 1,240,329 total parameters for the tested network. 
We designed the baseline to be as close in size to the baseline network as possible without shrinking the latent dimensions too quickly. 
Details of the convolutional network architecture are given in the appendix.

\subsection{Pre-Training Method}~\label{sec: pre_train}
The problem was difficult to effectively explore using $\epsilon$-greedy random exploration. 
To accelerate the learning process, we implemented a supervised pre-training step similar to the AlphaStar expert training~\cite{vinyals2019}. 
We used a stochastic rules based expert policy to generate a batch of example trajectories. 
The agent was then trained to minimize a composite loss term over this data. 

The pre-training loss term was based on a composite loss for distributional DQN~\cite{hester2018}.
The loss is defined as
\begin{align}
    \mathcal{L}_{P}(Q) = \mathcal{L}_{TD}(Q) +\lambda \mathcal{L}_{LM}(Q) 
\end{align}
where $\mathcal{L}_{TD}$ is the temporal difference loss defined in~\cref{eq: td_loss}, $\mathcal{L}_{LM}$ is a large-margin classification loss, and $\lambda$ is a weighting hyper-parameter.
The margin loss term is defined as 
\begin{align}~\label{eq: pre train}
    \mathcal{L}_{LM}(Q) &= \max_{a \in \mathcal{A}}\big[Q(h,a) + l(a_E,a)\big] - Q(h, a_E) \\
    l(a_E,a) &= 
    \begin{cases}
    Q(h, a) - Q(h, a_E) + \delta,& \text{if } a_E \neq a\\
    0,              & \text{otherwise}
    \end{cases}
\end{align}
where $a_E$ is the action selected by the expert policy and $\delta$ is the desired margin. 
This term encourages the values of actions taken by the expert policy to have a value higher than all others by at least $\delta$. 
Including the TD-loss term in the pre-training loss ensures that the value estimates are approximately Bellman-consistent, reducing forgetting early in RL learning. 

\section{Experiments}
We trained our neural network to solve the ACSO task using the proposed pre-training method and DQN algorithm.
Trained policy performance was evaluated over a batch of 100 episodes, with APT goal and attack vector fixed across episodes.
We ran experiments to evaluate the contribution of each novel element by comparing it to baseline policies and training methods. 
These experiments measured the performance of the converged policies and the sample efficiency of the learning. 

Both network architectures were pre-trained using the proposed composite loss.
These pre-trained networks were then tuned with RL for 500 episodes each. 
RL training was conducted using the hyper-parameters found during grid-search. 
Each architecture was also trained using DQN without pre-training, using 5 randomly initialized networks to test the impact of pre-training. 
Training details are in the appendix.
\begin{figure}[tb]
    \centering
    \begin{subfigure}[b]{0.49\columnwidth}
        \centering
        \includegraphics[width=\columnwidth]{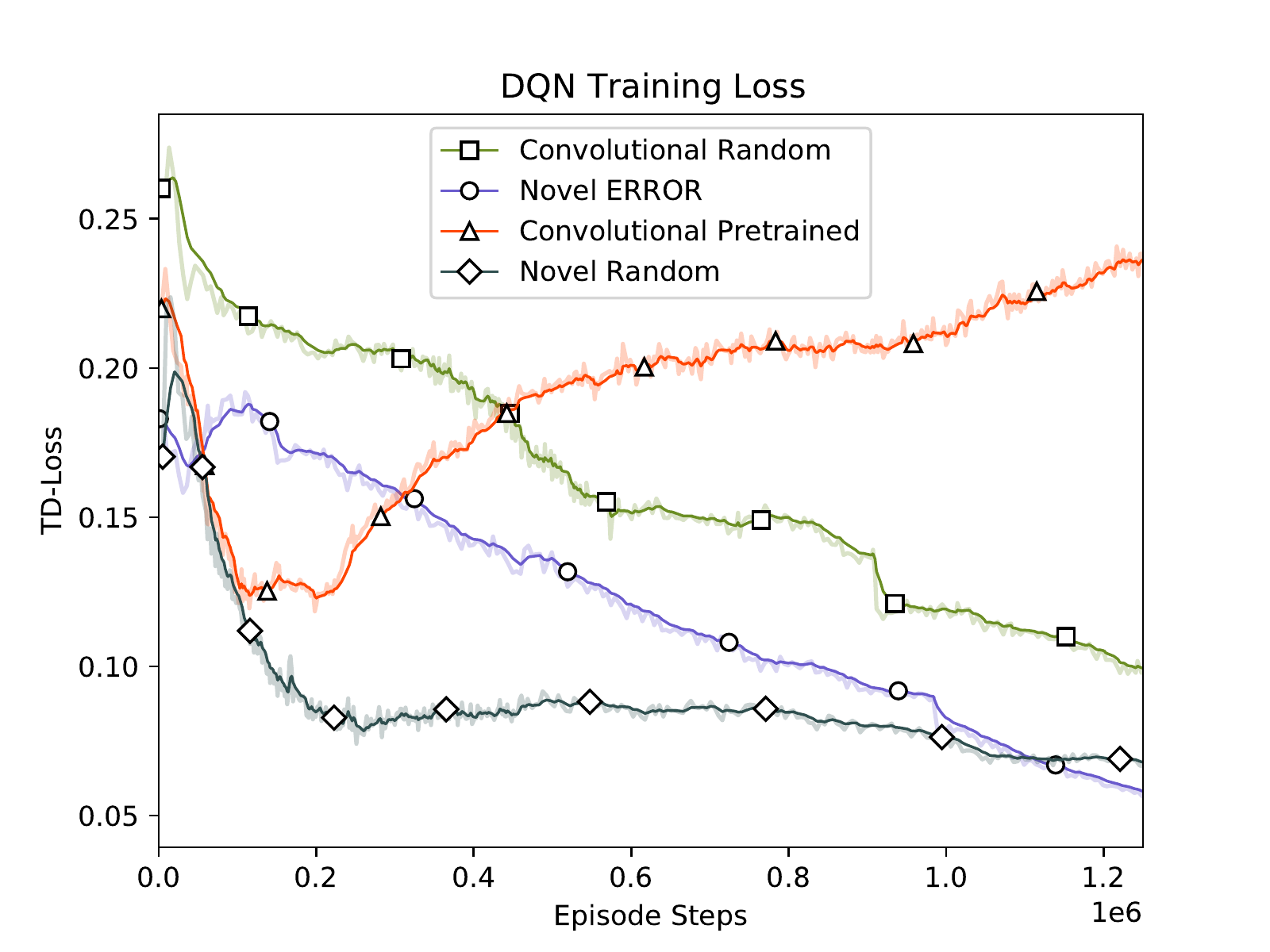}
        \caption{TD-Loss}
    \end{subfigure}
    \hfill
    \begin{subfigure}[b]{0.49\columnwidth}
        \centering
        \includegraphics[width=\columnwidth]{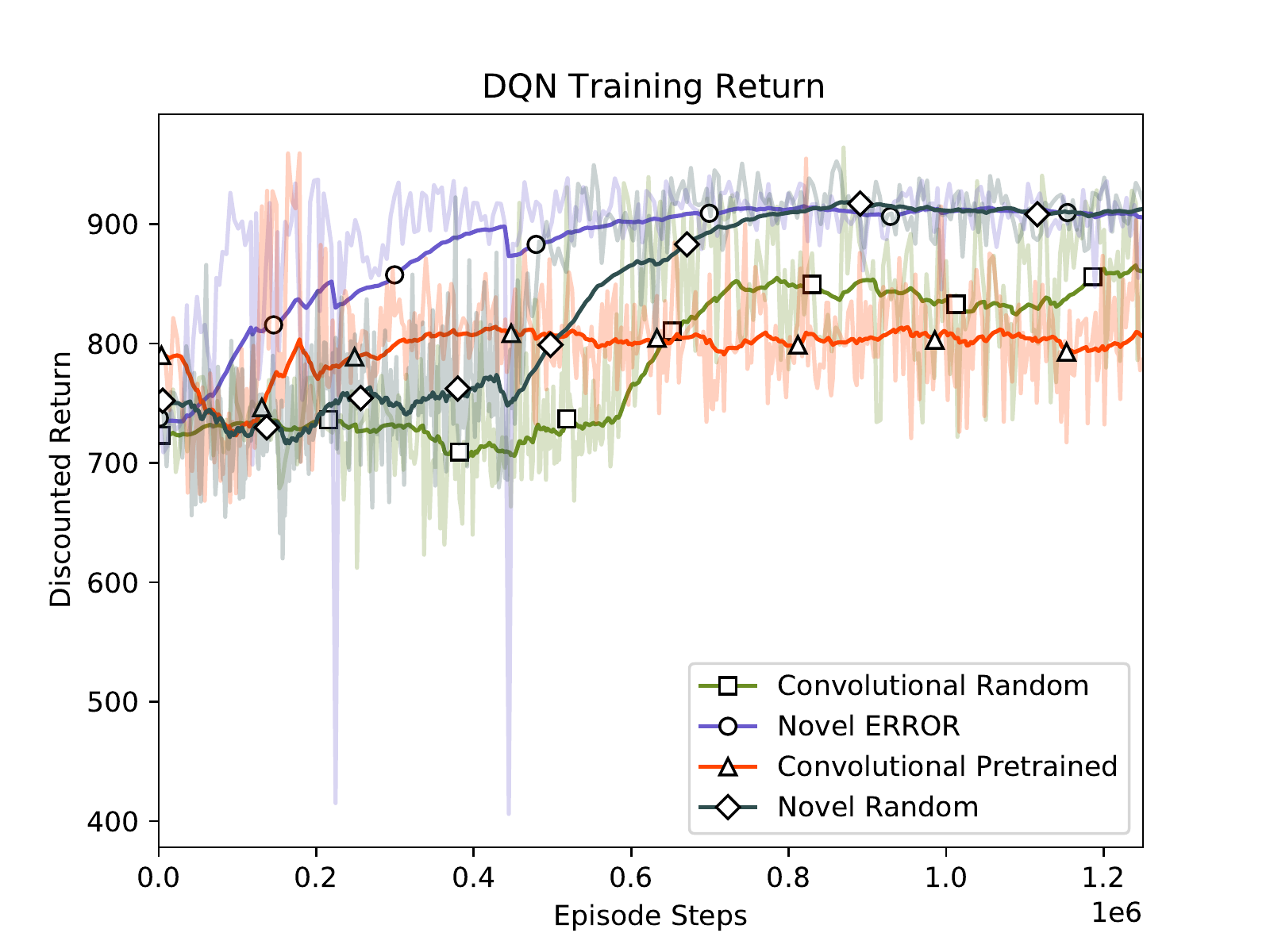}
        \caption{Discounted Return}
    \end{subfigure}
    \caption{Training Results: The left figure shows the average TD-loss per-episode for each architecture and pre-training combination. The right figure shows the time discounted sum of rewards for each episode. Both graphs show curves of the original data with an exponentially smoothed curve in bold.}
    \label{fig:learning_curves}
\end{figure}

\Cref{fig:learning_curves} shows the learning performance of each experiment configuration. 
For the randomly initialized networks, the best-performing seed is shown. 
The episode average TD-loss of~\cref{eq: td_loss} and discounted task return for 1.25 million episode steps are shown. 
As can be seen, our neural network TD loss decays much faster than the convolutional baseline both with and without pre-training, though pre-training tended to slow loss decay. 
The action-value margin of~\cref{eq: pre train} causes the agent to more often select actions similar to the pre-trained policy. 
This reduces how much the policy encounters other observation-action pairs, slowing learning of their values.
For the convolutional network, the TD-decay was initially accelerated with pre-training, however, the network quickly diverged after approximately 200,000 steps. 
This may be due to the the convolutional neural network being more sensitive to inconsistent action-value estimates induced by the pre-training.

The learning performance of the discounted return shows similar trends to the TD-loss, with our architecture generally reaching convergence faster the convolutional network. 
Additionally, our network's converged training performance was generally higher than the performance of the convolutional network. 
Unlike the TD-loss trends, pre-training accelerated the improvement in discounted return, allowing our architecture to approach converged performance much more quickly.
The pre-training negatively impacted the performance of the convolutional architecture. 

The evaluation performance of the best trained policy from each configuration is shown in~\cref{tab: performance}.
The mean performance over $100$ episodes along with one standard error bounds are given for four performance metrics. 
Task return is the time discounted sum of rewards defined by~\cref{eq: reward}.
The action cost is $10 \times$ the average cost per-step of defensive actions taken by the ACSO.
PLC downtime is the total number of hours that PLCs were off-nominal over each 5,000 hour episode.
Compromise time is the average number of IT nodes with an off-nominal state per simulation hour. 
The performance of the expert policy used to generate the pre-training data is also given.

\begin{table}[htb]
    \centering
    \begin{tabular}{l c c c c c} 
        \toprule
         Network & Pre-Train & Return & Action Cost & PLC Downtime & Compromise Time\\
         \midrule
         \multirow{2}{*}{Ours} & Yes & $940.4 \pm 0.0$ & $0.59 \pm 0.01$ & $\mathbf{0.0 \pm 0.0}$ & $\mathbf{0.2 \pm 0.0}$\\
          & No & $\mathbf{941.0 \pm 0.4}$ & $0.59 \pm 0.01$ & $\mathbf{0.0 \pm 0.0}$ & $1.0 \pm 0.0$\\
          \midrule
         \multirow{2}{*}{Conv} & Yes & $791.5 \pm 2.7$ & $\mathbf{0.26 \pm 0.01}$ & $6.2 \pm 0.2$ & $3.9 \pm 0.0$\\
         & No & $738.6 \pm 5.0$ & $0.57 \pm 0.01$ & $5.0 \pm 0.3$ & $3.5 \pm 0.1$\\
          \midrule
          Expert & -- & $906.9 \pm 4.5$ & $0.72 \pm 0.01$ & $0.2 \pm 0.0$ & $1.3 \pm 0.2$\\
         \bottomrule
    \end{tabular}
    \vspace{2mm}
    \caption{Evaluation Performance: This table shows the results of evaluating each trained policy over 100 trials. Evaluations of the expert policy used for pre-training trajectories is also given. The best performance for each metric is shown in bold.}
    \label{tab: performance}
\end{table}
As can be seen, our neural network greatly outperformed the convolutional network and the expert policy, both with and without pre-training. 
Our pre-trained network discounted return and action cost approximately equal those of the randomly initialized network.
Interestingly, the pre-trained policy had a significantly lower average number of compromised nodes per hour than any other policy, suggesting that it was able to benefit from the increased learning rate to find a more effective policy.
Both convolutional networks failed to outperform the expert policy, despite the expert policy trajectories having been used to pre-train the convolutional network. 
The novel networks outperformed the convolutional on task return despite having a higher action cost due to the fact that it successfully prevented more PLC downtime. 
This suggests that the actions it learned to take were both more aggressive and more effective at impeding APT progress.
Conversely, the expert policy incurred a higher average action cost than both novel networks. 
This suggests that though the expert policy was successful in preventing PLCs from going down, it was inefficient compared to the learned strategies.

\section{Conclusions}
This work presented a POMDP formulation of the cyber network security orchestration problem for ICS networks.
To enable this work and future research in this domain, we implemented a low computational cost ICS network attack simulation that is efficient enough for deep RL methods. 
A novel DQN architecture was shown to enable deep reinforcement learning to accommodate the large observation and action spaces of the problem. 
The proposed pre-training improved the learning performance of our architecture.

We ran several experiments to measure the effect each contribution had on learning efficiency and converged policy performance. 
The experiments show that both our proposed network architecture and the pre-training improve learning rate and converged performance.
The results suggest that deep reinforcement learning may be a viable method to implement automated security agents to mitigate the threat of APT attacks. 
The proposed contributions may also be applied to solve other large-scale, real world problems with discrete spaces. 
Results suggest that a neural architecture encoding the correct implicit bias can not only improve training efficiency, but may be required to learn an acceptable policy.

To build on this work, a simulation with increased fidelity to real-world network dynamics is needed.
Some future directions this may take are to tune parametric network models to real-world network data or to implement simulators with virtual machine kernels and real, known exploits. 
The former is likely the more promising path, as the latter poses significant computation and security challenges.
Including a model of an actual ICS process and linking the RL reward to process efficiency is also another path to investigate.
More robust attacker models with a wider set of attack vectors such as IDS signal spoofing or compromise of the ACSO compute assets should be investigated. 

More solution methods will also be explored with the proposed network architecture.
Recurrent neural networks and variational networks are alternatives to sliding window histories that may provide better beliefs over the network state.
Integrating attention mechanisms with graph neural networks~\cite{velickovic2018} may enable application of learned policies to ICS networks with dissimilar architectures.
Policy gradient methods, such as phasic policy gradient~\cite{cobbe2020}, will also be explored as an alternative to DQN. 

The current work only considers a fixed, stochastic attacker model.
To improve the robustness of learned policies, adversarial learning will be explored to generate more challenging attacker strategies. 
Similarly, the alert and transition dynamics were modeled as stationary, though these may evolve in real-world network operations.
Stress testing of these dynamics will also be considered.
Assurance and verification of ACSO actions, especially on safety-critical ICS will be investigated.

\begin{ack}
This material is based upon work supported by the Johns Hopkins University Applied Physics Laboratory.
The work was supported by the Stanford University Institute for Human-Centered AI (HAI) and Google Cloud. 

\end{ack}


\printbibliography

\end{document}